\documentclass[11pt]{article}
\usepackage[margin=1.0in]{geometry}
\usepackage{graphicx}
\usepackage{epsfig}
\usepackage{amssymb,amsfonts}
\usepackage{epstopdf}
\usepackage{hyperref,color}
\usepackage{natbib}
\usepackage{setspace}

\begin{document}

\title{Emergent Spacetime and Empirical (In)coherence}
\author{Nick Huggett and Christian W\"uthrich\thanks{Work on this project has been supported in part by a Collaborative Research Fellowship by the American Council of Learned Societies.}}
\date{}
\maketitle

\begin{abstract}\noindent
Numerous approaches to a quantum theory of gravity posit fundamental ontologies that exclude spacetime, either partially or wholly. This situation raises deep questions about how such theories could relate to the empirical realm, since arguably only entities localized in spacetime can ever be observed. Are such entities even possible in a theory without fundamental spacetime? How might they be derived, formally speaking? Moreover, since by assumption the fundamental entities can't be smaller than the derived (since relative size is a spatiotemporal notion) and so can't `compose' them in any ordinary sense, would a formal derivation actually show the physical reality of localized entities? We address these questions via a survey of a range of theories of quantum gravity, and generally sketch how they may be answered positively. 
\end{abstract}


\section{Local beables and empirical coherence}

A central concern of philosophy of science is understanding how the theoretical connects to the empirical, the nature and significance of `saving the phenomena'. This is not the place to propose another theory describing, or prescribing, this connection; let alone to consider how such a theory might, in turn, relate to how science actually works. At a high level of generality, however, presumably the link is established by observing (in some sense) a material `something', in some determinate state or other, at some spatial location at some moment in time and connecting this occurrence to our theory, for instance by postulating, in our theory, entities which behave in ways that would explain our observation. This is crude, no doubt, but seems to capture quite generally the nexus between our theorizing about the world and our experiencing it, from meter readings in the lab to observing distant galaxies with a radio telescope to the results of high energy collisions.

The bottom line is that empirical science as we understand it presupposes the existence of what John \citet[234]{bel87} called `local beables'. `Beables' are things that we take to be real, from fundamental objects of the theory to more familiar objects of experience (in the context of interpreting quantum mechanics they are candidates for being---hence they are `be-able'). `Local' is a term with a number of meanings, some technical, but Bell has in mind locality in the sense of being ``definitely associated with particular space-time regions''---so `local' here does not carry any direct consequences for interactions or their propagation. What he has in mind is closely related to the principle of `separability' proposed by Albert \citet[\S2]{ein48}. For our purposes it is generally adequate to take a beable to be local if the degrees of freedom describing it are associated with an open region of spacetime. It should be noted that this condition for locality is very weak as it stands---entities spreading across different galaxy clusters still qualify as `local'. Hence, we think of this locality condition as necessary, but likely not sufficient, for observables. In at least one of the cases discussed below, the notion will be tightened. 

Local beables then represent the material content of our universe, ultimately constituting what we observe, though of course most specific properties postulated of these local beables depend on the specific theory at stake. At the fullest level of generality, however, they share one crucial feature: as Tim \citet[3157]{mau07} puts it, ``local beables do not merely exist: they exist somewhere.'' 

Maudlin's paper is a criticism of `configuration space' interpretations of quantum mechanics, but it ties together the 
ideas just developed to challenge any theory (plus interpretation) which claims that fundamentally, familiar space and time do not exist. `Normal' theories, which postulate space and time, allow for local beables in their fundamental ontology, so the observable local beables can potentially be understood as composed, spatiotemporally, of the fundamental ones. A theory that does not postulate familiar space and time in its fundamental ontology precludes fundamental local beables, and there is then no obvious strategy for identifying observable local beables---and the question of the theory's empirical significance becomes acute. Indeed, the problem is not merely one of not knowing how to test such theories. Maudlin argues that a mere technical solution will not suffice---a claim we shall address in in the final section of the paper. For now consider another problem (first raised by Richard \citet{hea02} in a more limited context).

Following Jeff \citet[\S4.5.2]{bar99},\footnote{Cf.\ also \citet[50]{bar96}.} we define a theory to be {\em empirically incoherent} in case the truth of the theory undermines our empirical justification for believing it to be true. Thus, goes the worry, if a theory rejects the fundamental existence of spacetime, it is threatened with empirical incoherence because it entails that there are, fundamentally, no local beables situated in spacetime; but since any observations are of  local beables, doesn't it then follow that none of our supposed observations are anything of the kind? The only escape would be if spacetime were in some way derived or (to use the term in a very general sense, as physicists do) `emergent' from the theory. But the problem is that without fundamental spacetime, it is very hard to see how familiar space and time and the attendant notion of locality could emerge in some way \dots\ at least without some concrete proposals on the table. 

\citet{hea02} similarly argued that the absence of time (or at least of some of its essential features) in a fundamental theory threatens the theory's empirical coherence in a way that can only be averted by establishing ``how we can get along with a concept of time that lacks that feature." (293f) He is only concerned with the challenges raised by the `problem of time' in canonical quantum gravity \citep[\S2]{hugeal12}, but the general point clearly applies to the `problem of spacetime' in quantum gravity generally. In the present paper we explore how, extending his treatment to new cases, and sketching answers to the questions he raises.

Suppose then that as far as many quantum theories of gravity are concerned, in various ways, familiar spacetime is not admitted at the fundamental level, putting quantum gravity in violation of Maudlin's dictum, and threatening empirical incoherence. In the next section, we will consider a range of such theories, and observe that the seriousness of these challenges depends a great deal on what they postulate instead of spacetime. Different theories leave more or less of the standard structure of spacetime intact, and so understanding our observations may, in the best case, require only a relatively small shift in our conception of local beables. But we will also see that in theories in which little or nothing of spacetime is left in the fundamental ontology, it still may be the case that the question of deriving some formal structure that mirrors local beables can be answered rather more readily than one might expect. In the final section we will turn to such derivations and address Maudlin's argument that such formal derivations never show that local beables are part of an emergent ontology.

\section{Theories without spacetime}

Quantum gravity's denial of the `spacetimehood' of the fundamental structure thus comes in degrees. Some theories of quantum gravity characterize this structure only as somewhat different from relativistic spacetimes, with at least apparently straightforward connections between the two conceptualizations, while others conceive of it in radically novel ways, bearing hardly any resemblance to relativistic spacetimes. We thus find a spectrum of theories with increasingly iconoclastic re-conceptions of `spacetime'. It is important to pose the challenge of empirical incoherence to them separately: for as we shall show, they face greater or lesser challenges, of different kinds, in finding spacetime, and hence with empirical coherence.\footnote{The cases we consider overlap partially with those discussed in \cite{D.M:05}. While we stress the ways in which spacetime `fades' away, their complimentary work studies senses in which such theories remain geometrical.}

However, before proceeding, we note that there really is no single `spectrum' of cases, linearly ordered by some single factor encapsulating the `spacetiminess' of the structures in question. Instead, one should picture the situation as more akin to an at best partially ordered field of theories in a space with multiple dimensions corresponding to ways in which these structures depart from relativistic spacetimes. In particular, the way one might rank theories conceptually, according to how much of the concept of spacetime they maintain, does not always align with the ranking according to the ease with which putative local beables or their surrogates can be found. Mainly for reasons of presentation, our list will more closely follow the former scale; but bear in mind that the picture we are presently drawing is only a qualitatively valid representation of the situation as we see it. We now proceed to describe six types of departures from relativistic spacetimes, loosely ordered in decreasing similarity to relativistic spacetimes.

\subsection {Lattice spacetime} 

As a first step away from familiar relativistic spacetimes, we find discrete lattices. Many expect the smooth classical spacetime to be coarsened into a discrete structure at the quantum level. For instance, Lee \citet[549]{smo09} takes discreteness to be ``well established'' and a ``generic consequence'' of a ``large class'' of quantum theories of gravity. Four statements characterize the class of theories Smolin is interested in. First, the usual basic postulates of quantum mechanics are assumed, although Smolin does not elaborate whether this includes, e.g., a form of a collapse or projection postulate. Second, at least partial background independence is presupposed in the sense that theory makes no reference to a fixed spacetime background, even though some features of the fundamental structure such as its dimension and topology may be fixed. Third, the fundamental structure is stipulated to be a set of basal elements partially ordered by causality. Finally, what Smolin confusingly labels ``discreteness'' is assumed, viz.\ that the Hilbert space of the quantum theory is separable, i.e., has a countable basis of combinatorial structures. The dynamics of the theory arises from `moves' which are local in the topology of the fundamental structure. Examples of theories which fall within the purview of these assumptions, according to Smolin, are causal set theory, dynamical triangulations, consistent discretization models and quantum causal history models. (For reasons explicated below, we take some of these theories, notably causal set theory, to be of type (ii), to be discussed shortly.)

It may be complained that Smolin regards as a substantive consequence something that theories satisfying these assumptions must already conform to. As the fourth postulate requires a form of discreteness to hold of the theories at stake, it should come as no surprise that these theories are ``generically'' discrete! There is something to this complaint, although it should be made clear that the discreteness assumed and the discreteness derived are physically inequivalent statements: while the fourth postulate requires that the basis of the Hilbert space is a countable set of combinatorial basal elements, the consequence, which can be rigorously established, is that theories satisfying the assumptions are ultraviolet finite. No doubt this consequence is of great physical significance. What matters for our purposes is not the ultraviolet finiteness in itself, but yet another form of `discreteness', viz.\ the discreteness of the geometry of the fundamental structure, for instance in the form of an indivisible shortest length, which of course delivers the cut-off necessary to render the theory ultraviolet finite. The relevant form of discreteness may be postulated {\em ab initio}, as it is in causal set theory, or be the consequence of the axioms of the theory which do not directly postulate discreteness, such as arguably is the case in loop quantum gravity. But whichever notion we take to be most descriptive of `discreteness', whether discreteness (or some other feature of the fundamental structure such as non-commutativity) is {\em postulated} or {\em inferred} is immaterial for our present purposes. What matters, ultimately, is the nature of the fundamental structure and how it differs from general-relativistic spacetime. 

One step removed from relativistic spacetimes, then, we find discrete lattices consisting of a set of basal events exemplifying a structure spanned by the spatiotemporal relations obtaining among pairs of such events. At this first step, it is important that the relevant relations be interpreted spatiotemporally, in such a way that the only relevant difference is that what was a continuum before is now a lattice. To facilitate the discussion, let us take the image of a lattice somewhat seriously and imagine, without much loss of relevant generality, that all spatiotemporal relations among the lattice points supervene on a basis of elementary spatiotemporal relations capturing spatiotemporal adjacency. We can thus think of these elementary relations as `spacing' the basal events---the nodes of the lattice---equidistantly in spatiotemporal terms. The structure thus possesses a minimum spatiotemporal distance. The discreteness of the structure is captured by the demand that any two basal events can be connected by elementary relations with only finitely many intermediate events. 

The situation is thus analogous to the transition from continuous matter to atomism. One can (almost) imagine a seventeenth-century Maudlin, arguing that since all our observations are of continuous bodies, our fundamental theories must postulate their existence (some Cartesian arguments have this flavour). But it is not too hard to think of ways in which atomic matter might appear continuous. Perhaps the interstices might not be observable to the naked eye---because we lack the power to resolve the small variations involved. Similarly, the twenty-first-century Maudlin's challenge doesn't get much bite (nor is it obviously intended to) on theories of discrete spatiotemporal lattices. Without downplaying the challenges arising from understanding this emergence of a continuum from a lattice structure, it seems clear that the discreteness of the structure that physics has hitherto taken to be (smooth) spacetime alone does not undermine the localization required for empirical coherence. The local beables would presumably simply sit at the nodes of the lattice, either each at a single node or distributed over a set of nodes which was simply connected in terms of the adjacency relations in the supervenience basis. 

If spatiotemporal discreteness were thus the only prediction made by most, or at least many, quantum theories of gravity, the challenge to their empirical coherence would seem easy to defeat. We insist, however, that even this comparatively mild divergence from relativistic spacetime would raise important questions regarding the emergence of the continuum from the lattice and regarding classical limits of the theory. The trouble is that quantum gravity may necessitate more radical departures.

\subsection{Non-metrical lattices}

Moving only slightly further away from relativistic spacetimes we find approaches to quantum gravity that also describe the fundamental structure as discrete, but where the relations connecting the basal events are not the ordinary spatiotemporal ones. For instance, causal set theory assumes two basic kinematic axioms, viz.\ that an elementary relation of \emph{causality} imposes a partial ordering on the basal events, and that the structure is discrete in the sense explicated above. The approach derives its motivation from results in general relativity, such as a theorem due to David \citet{mal77}, which establish that the causal structure of spacetime is enough to recover almost everything else that matters, most importantly the metric relations obtaining among the spacetime points, up to a conformal factor. These results do not mean that relativistic spacetimes can be recovered from structures satisfying the kinematic axioms alone---the required discreteness `thinned out' the structure too much, as it were. The hope of proponents of this program is that additional dynamical constraints will lead to a principled story of how these fundamental causal structures give rise to the much richer structures of relativistic spacetimes. 

The elementary causal relation departs from ordinary spatiotemporal relations in three respects. The first significant difference is that the fundamental relation in causal set theory does not wear a metric on its sleeve. There simply is nothing on the fundamental level corresponding to lengths and durations (or, more generally, to spacetime intervals), and no alternative interpretation of the causal sets in terms of metrical relations is available. The second deep difference is that the theory lacks the structure to identify `space', in the sense of a spacelike hypersurface. One can of course find sets of basal elements such that no pair exemplifies the causal relation, but such `antichains' have no structure beyond their cardinality. In other words, nothing but a difference analogous to that between spacelike and timelike remains at the fundamental level. Finally, there is a sense in which the discreteness of the causal set and the Lorentz invariance demanded of the emerging spacetime conspire to render the physics non-local in a way unfamiliar to general relativity.\footnote{Cf.\ \citet{sor09}.} Hence, it is much more correct to conceive of the fundamental relation as {\em sui generis} causal, and not, strictly speaking, spatiotemporal. Spacetime, together with spatiotemporal relations, emerges from the fundamental structure.

The work that needs to be done to recover the spacetime structure from the causal set in some continuum limit thus transcends that which was necessary in the simple lattice case. The additional work arises because the fundamental structure cannot directly be interpreted spatiotemporally. Even though in general relativity it may be a small step from the causal structure to the spatiotemporal, the discreteness of causal sets interferes with such a direct line of reasoning. As it turns out, almost all causal sets satisfying the above kinematic axioms fail to have well-behaved continuum limits resembling relativistic spacetimes.\footnote{This is what \citet[\S7.5.1]{smo06} calls the `inverse problem'.} Thus, the underlying discrete causal structure is at least sufficiently different from spacetime that this task becomes significant. 

However, it does not immediately follow that we have finally reached a case in which empirical incoherence becomes a significant worry. Even if it is hard to find the {\em spatiotemporally local} beables necessary for empirical science, another strategy suggests itself: take localization in causal terms, and argue that it is causal nexus, rather than spatiotemporally understood locality, which supplies the  condition relevant for empirical coherence. We won't pursue this proposal, but we do want to say that we have reached a transitional case of a theory in which the weakening of familiar spacetime structures makes the derivation of local beables at least a challenging question.\footnote{For a more detailed philosophical analysis of causal set theory and its challenges, cf.\ \citet{wut12}.}

\subsection {Loop quantum gravity}

 Somewhere even further removed from relativistic spacetimes we find the structures described by loop quantum gravity (LQG), one of the main contenders for a quantum theory of gravity. LQG starts out from a Hamiltonian formulation of general relativity, which is then subjected to a canonical quantization procedure. This procedure has not been completely executed, and therefore what follows here remains tentative. There are two important ways in which the loop structures of LQG differ from a discrete lattice so as to further diverge from relativistic spacetimes, both having to do with the quantum nature of the fundamental structure. In order to appreciate them, let us say a few words characterizing this structure. 

What is believed to correspond to three-dimensional {\em spatial} structures are so-called `spin networks'. Spin networks can be thought of as networks of interwoven loops with `spin' representations sitting on both the network's nodes and its edges. These spin representations quantify the discretely valued quantum `volume' to which the node corresponds, and the discretely valued quantum `area' of the edge corresponding to the surface of adjacency of the connected `volumes'.\footnote{These spin representations take as values the eigenvalues of quantum operators for which the relevant spin network states are eigenstates. These operators afford a natural geometric interpretation, as indicated in the main text, but they are not Dirac observables, i.e., they are not gauge-invariant. It thus remains an open question to what extent the offered physical interpretation of these states will be vindicated in the final form of the theory.} Although the dynamics of the theory is far from settled, the general scheme evolves these spin network states by having an appropriate Hamiltonian operator acting on them. In general terms, the action of the Hamiltonian on a node of the spin network will either be identity, i.e.\ the node simply persists through `time', or it splits the node into several nodes, or it fuses the node and some other nodes into a single node. The resulting structure is taken to be the quantum analogue of a four-dimensional spacetime and is called `spin foam'. As the dynamics of the theory remains ill-understood and we have greater technical control over the Hilbert space of spin networks, we will proceed, without loss to the pertinent philosophical points, with our analysis by focussing on spin networks. 

The spin networks bear a superficial resemblance to the discrete lattices introduced above, but there are two relevant differences. First, the actually existing and physically fundamental structure is supposed to be a quantum superposition of something like these spin networks, and not just a single spin network. Since all the different structures in the superposition will have a different connectivity (and perhaps different cardinality), and in this mathematical sense be different structures altogether, what is local in one term of the superposition will in general not be local in others. Except perhaps for very special states, local beables can thus not be part of the fundamental reality, but must instead emerge in some limit---presumably the same as that in which locality emerges. How such local, i.e.\ topological, structures like relativistic spacetimes emerge from spin networks is at present little understood. 

\begin{figure}
\centering
\epsfig{figure=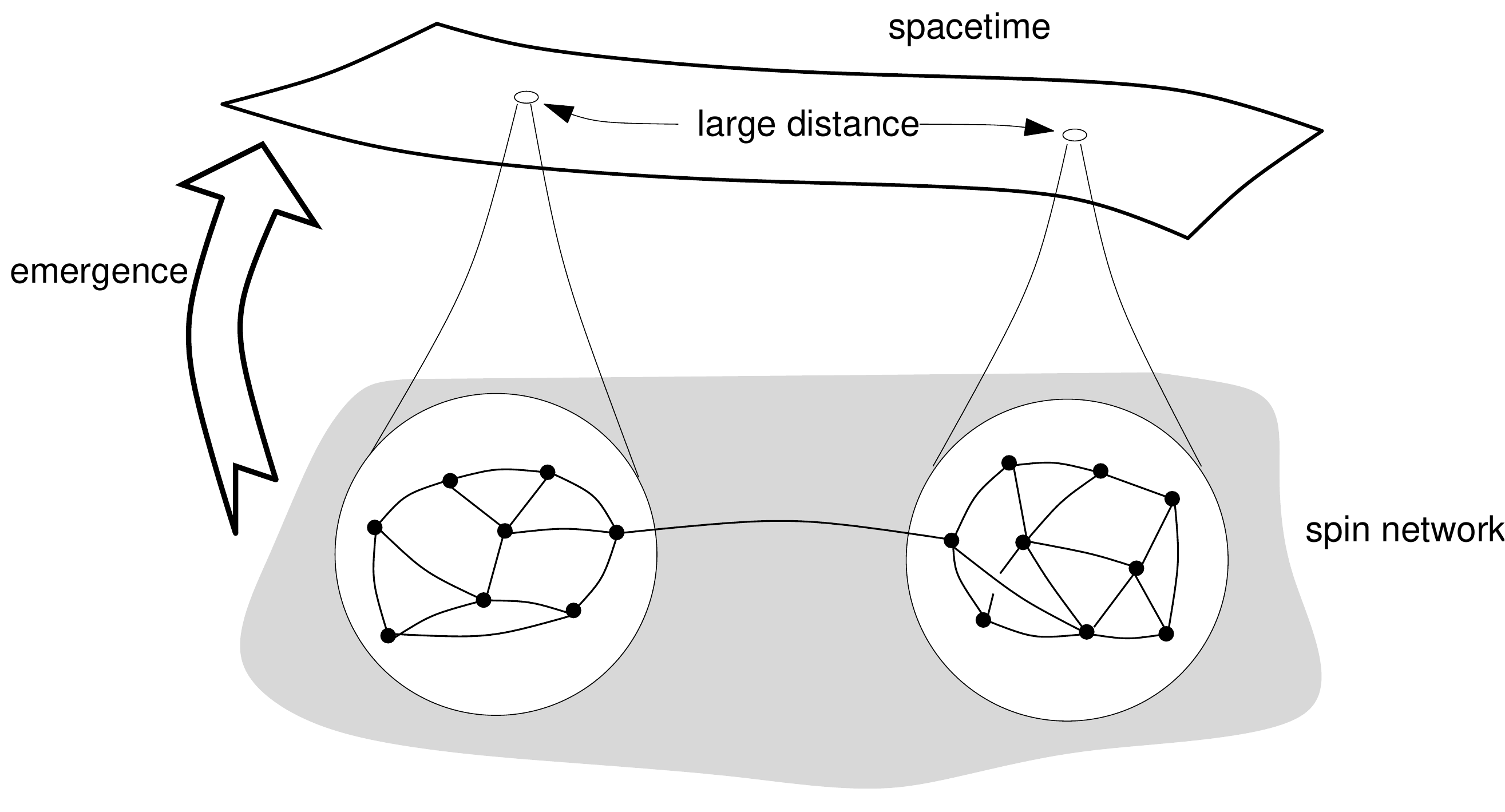,width=\linewidth}
\caption{\label{fig:nonlocality} A spin network (shown as shaded) with two regions (shown enlarged) connected by an adjacency relation and how they are embedded in an emerging spacetime.}
\end{figure}

Secondly, not only does the quantum superposition frustrate the applicability of locality criteria, but there is a sense in which even a spin network corresponding to a single term in the superposition is not amenable to the kind of localization that may be required to ensure empirical coherence. The problem is that any natural notion of locality in LQG---one explicated in terms of the adjacency relations encoded in the fundamental structure---is at odds with locality in the emerging spacetime. In general, two fundamentally adjacent nodes will not map to the same neighbourhood of the emerging spacetime, as illustrated in Figure \ref{fig:nonlocality}. Hence the empirically relevant kind of locality cannot be had directly from the fundamental level. 

Let us explain this in slightly more detail. In order to appreciate the point, how are the fundamental spin networks supposed to be related to relativistic spacetimes? This is a vexed issue which so far defies resolution---in fact, it is the hardest problem tormenting loop quantum gravitists. One influential idea based on so-called `weave states' proposes that the spacetime structure emerges from appropriately benign, i.e.\ semi-classical, spin networks which are in eigenstates of area and volume operators (or perhaps other geometric operators), with eigenvalues which approximate, for sufficiently large `chunks' of the spin network, the classical values of the standard area and volume functions of the corresponding approximated chunk of spacetime.\footnote{For more details on this idea and on the difference and relation between relativistic spacetimes and the fundamental structures posited by LQG, cf.\ \citet{wut12b}. For a more comprehensive discussion of semi-classical strategies, cf.\ \citet[\S11]{thi07}.} 

Locality, or vicinity, in the emerging spacetime is determined by the spatiotemporal---indeed metric---relations obtaining between the events of spacetime. `Locality' is notoriously hard to capture in general relativity, but if you are bothered by our loose use of the term here, imagine, for what follows, that the relativistic spacetimes of concern are simply the standard FLRW spacetimes, which admit a preferred foliation into `spaces' of constant spatial curvature, linearly ordered by `cosmological time'. `Local' could then be understood as `nearby' as measured in the induced metric of these privileged `spaces', where `nearby' can be specified in the experimental context of concern. Note that this notion of locality is strictly stronger than the one proposed as necessary, but perhaps not sufficient, at the outset: there, we only demanded that there be an open region of spacetime supporting the local beable, and we placed no restriction on the size of the region. Here, this condition is strengthened to require that these open regions supporting local beables do not exceed some contextually specified spatial volume.

Adjacency, the corresponding {\em fundamental} relation, is captured by the combinatorial connectivity of the spin network structure. Suppose we have isolated the way in which a relativistic spacetime emerges from the spin network, and thus how the basal events of the spin network get mapped into the spacetime. In general---and here is the crux---pairs of adjacent basal events need not be mapped into Planck-sized neighborhoods of one another in the emerging spacetime: as judged in the metric of the spacetime, these events are thus not `nearby'. For any given maximum spatial volume of the locality-supporting open region (Planck-sized or more), there will in general be some adjacent pairs of basal events one which gets mapped into the open region, and the other one outside of it. Now, these `non-localities' are suppressed in the approximation of the large scales relevant to the emergence of spacetime. A particular spacetime would simply not emerge from the fundamental structure unless the adjacencies it ignores are `weak', i.e., washed out by those adjacencies it respects. If the adjacencies are of sufficient strength (as defined by their spin representation) and number between groups of events, then only a spacetime which places them near one another could emerge from the corresponding spin network, at the expense of yet other, weaker, adjacencies.

As a consequence of the two distinct ways in which locality is threatened in LQG, establishing the emergence of relativistic spacetimes, based e.g.\ on the weave-states idea, will presumably involve two distinct kinds of technical procedures. The first procedure, an {\em approximation} in the sense of \citet{butish99,butish01}, should show how the dynamics forces the quantum state into semi-classical states with a well-behaved classical counterpart such that, e.g., the quantum superposition is dominated by a single spin network. The second, {\em limiting}, procedure then establishes the connection from the semi-classical states to classical relativistic spacetime. A thorough analysis of these procedures in the context of LQG will have to await another day, but it should be clear that significant work is required to regain the original and arguably pertinent sense of locality.\footnote{For a bit more on how spacetimes may emerge from LQG, cf., again, \citet{wut12b}.} 

The entire programme of LQG stands and falls with the success of this endeavour: if we fail to recover relativistic spacetimes from the spin foams and spin networks of LQG, and hence fail to understand the relationship between general relativity and LQG, there is no prospect of explaining why general relativity was as empirically successful as in fact it was while at the same time false, and therefore in need of replacement. However, it should be noted that {\em every} theory hoping to dislocate a fundamental theory which is `incumbent' at the time must discharge this explanatory debt. We see no other way for LQG of doing so unless it is understood how spacetimes emerge from spin networks. Once this {\em is} understood, however, the threat to LQG's empirical coherence is {\em ipso facto} averted.

\subsection {String dualities}

The next two examples that we will discuss are drawn, not from attempts to quantize general relativity, but from attempts to incorporate gravity within QFT, and in particular from the string theory program. It might seem that string theory is not a very interesting case for discussion here, because on a naive reading it looks exactly as if strings are local beables, bits of stuff describing worldsheets in a classical spacetime. However, the discovery or intuition of various `dualities' complicates this verdict.

To say that two theories are `dual' is to assert that they are in some sense equivalent descriptions: under some mapping, the relevant content is preserved. That general characterization leaves a lot open of course: what constitutes a theory? What exactly should be preserved? What kinds of mappings are allowed? Rather than trying to answer those general questions, we will explain the idea with an example: quantum harmonic oscillators are dual under the mapping $\langle q, p; m, k\rangle \to \langle p, -q; 1/k, 1/m\rangle$ (where $m$ is the mass and $k$ the spring constant). The duality exists because there are contributions to the energy from both position ($q$) and momentum ($p$), in such a way that the form of the Hamiltonian is preserved by the mapping, along with the canonical commutation relations: hence the map preserves the energy spectrum and the expectation values of dual observables. Of course, as theories of particular quantum oscillators, these are not equivalent theories at all, for there is a difference between mass  $m$ and mass $1/k$, and there is a difference between the state being $\psi(q)$ and its being $\psi(p)$ (in which case the spatial wavefunction is the Fourier transform, $\tilde\psi(q)$). That is, one finds physical differences when one steps back from the theory and looks `from the outside', to give significance to mass, position etc. Things are different in string theory because it is a `total theory', for which there is no broader context in which to distinguish duals in this way.

Two so-called `T-dual' string theories live in compactified spaces---one or more spatial dimension has the topology of the circle---in which the radius of a compact dimension in one is (in suitable units) the reciprocal of the radius in the other. There are contributions to the Hamiltonian from the kinetic energy of the string, which decreases with wavelength; and from energy due to tension in the string, which increases with its length, so with the number of times the string winds around the dimension. Call this latter the `winding number'. Thus for fixed wave-number and winding-number around a circular dimension, as its radius grows the kinetic energy decreases while the energy due to winding increases. Hence, for suitably related radii, the Hamiltonian is preserved under the interchange of wave- and winding-numbers: specifically, the dual radii are reciprocal. To quantize winding, the integer spectrum requires the string to have a winding state described by a wavefunction around a new closed dimension parameterized by a continuous coordinate, $w$: the winding number is represented by the wave number of $\psi(w)$. As \citet[29]{wit96} puts it, the $w$-coordinate describes ``another `direction' peculiar to string theory''. In this representation, the duality map exchanges the position and winding degrees of freedom: $\langle q, w; R\rangle \to \langle w, q; 1/R\rangle$. As with any `position' there is a momentum conjugate to $w$, and as long as the mapping is understood also to exchange ordinary and winding momenta, it preserves canonical commutation relations as well as the Hamiltonian, and we have dual theories just as before.

The formal story may be much the same in the two cases, but now that we have a total theory, the interpretive consequences are quite different. There are no physical quantities, processes or entities beyond those of the theory itself that might determine whether the radius of the compact dimension is $R$ or $1/R$, and all we can do is ask what the theory says: but we know that whatever value one theory assigns to a quantity, the dual theory assigns the same value to the dual quantity. That is, on the assumption that the observables of the theory exhaust its physical content, there is no physical difference which could settle the issue: in particular, the result of any possible observation, using any technology whatsoever permitted in theory, has equally valid but distinct interpretations in the two theories.\footnote{To give a concrete illustration, \cite{BraVaf:89} consider timing a photon around a closed dimension of phenomenal space to determine its circumference. Suppose the result is $2\pi R$. One model of the situation has a string living in a space with a compact dimension of radius $R$, and with spatial wavefunction $\psi(q)$. In the dual model the compact dimension has  radius $1/R$, but the photon has the dual state $\psi(w)$; this state evolves dually along the winding coordinate, in effect describing a photon `moving along' that `direction' (further illuminating Witten's point). It turns out (unsurprisingly) that in this dual model the winding coordinate has radius $R$, so the photon takes the same time to `travel around' the winding coordinate as the original photon did to travel around space. In this dual theory then, our phenomenal space does not correspond to the space of the string, but to the corresponding winding dimension.} Since the theory with radius  $R$ is physically equivalent to the theory with radius $1/R$, the difference between the possible radii of the space in which the string lives is not physical, but surplus representational structure. The two theories are, after all, merely two representations of the same physics, and any differences are unphysical: thus the radius of the space in which the string lives according to string theory (i.e., in either representation) is not physically determinate.

But of course a compact dimension in phenomenal space does have a determinate physical radius: if one of the three familiar spatial dimensions turns out to be compact, it has a very large, not very small radius;\footnote{T-duality proofs generally assume a locally Minkowski geometry, and there seem to be differing opinions about whether the duality would also hold in the more complex geometry of the familiar dimensions.} and if space turns out to be Kaluza-Klein then the extra dimensions are very small, not very large. Therefore, \emph{the space in which a string lives is not phenomenal space}: the former does not have a determinate radius, while the latter does. (Unless one puts one's foot down and proclaims a real but otiose fact of the matter about which of the dual string theories is correct, but that is to introduce `hidden variables' into the theory, for no clear reason.\footnote{For a more principled introduction of a hidden radius see \cite{Nik:07}.}) String theorists generally conclude that there is some underlying `M-theory' in which these differences are clearly merely unphysical choices of representation, and from which phenomenal space emerges. However, the threat of empirical incoherence is clear in string theory: since strings aren't simply objects in phenomenal space, how do they (how could they) relate to the local beables which ground experiment?

Other dualities, demonstrated or postulated, put similar pressure on even more fundamental (pre-metrical) elements of spacetime structure. Mirror symmetry \citep{Ric:10} relates spaces of different topologies, not just metrics, and the AdS/CFT correspondence (Rickles, Teh, this issue) relates spaces with different dimensions. The lesson seems the same: none of these properties can be taken as physically determinate in the fundamental picture of the world in string theory.\footnote{One might argue that the difference between T-dual theories is not representational, but merely notational: since the $q$- and $w$-dimensions exchange radii, one might say that T-duality simply relabels them. Then the spatial dimension would be that whose radius matches that of phenomenal space, whether parameterized by $q$ or $w$: the apparent indeterminacy of radius merely reflects the conventional choice of labels. The additional dualities offer even more convincing cases of spatial indeterminacy because the relate theories that cannot be similarly understood as notational variants.} But again, if strings live in spaces without these features then strings are not local beables in phenomenal space (and neither are D-branes, for the same reasons). However its lack of local beables does not cause a problem of empirical incoherence for string theory, because in each case one of the duals has a spacetime matching phenomenal spacetime (e.g., in one of the T-dual theories the $q$-radius equals the phenomenal radius)---observables that are local in that representation correspond to local phenomenal quantities. For the reasons given above, to say the spaces `match' is not to say that space in one of the duals \emph{is} phenomenal space, but simply that the correspondence provides a way to read off observable quantities corresponding to familiar spacetime phenomena, thereby deriving empirical predictions. Neither are the fundamental objects of that dual theory local beables in phenomenal spacetime: for instance, there is no physical fact of the matter how long strings are, or how many times they are wrapped around space, whereas (by definition) these properties are perfectly determinate for local beables in familiar spacetime. (To make the point one last time, such string properties are not physical, but surplus representational structure.) Nevertheless, though dualities eliminate local beables from the basic furniture of the world, they do not introduce any technical problems for deriving predictions in familiar spacetime that would entail empirical incoherence. The correspondence between one dual and phenomenal spacetime of course explains how it is that string theorists make predictions, for instance about string scattering or the entropy of black holes. However, in the final section we will return to the question of whether there are conceptual problems concerning the kind of derivation of predictions sketched here.

\subsection {Non-commutative geometries} 

It turns out that certain string theories have a `non-commutative field theory' (NCFT) in their low energy limits (e.g., \cite{seiwit:99a}). These are field theories living in a spacetime in which the coordinates do not commute: literally, $x\cdot y\neq y\cdot x$. What does this mean? Well, the connection between algebra and geometry has of course been known a long time. Beyond the kind of relation described by Descartes, the differential geometry of a manifold can be uniquely characterized by the algebra of scalar and constant fields on it. For instance, consider the algebra of fields polynomial in commuting coordinates: $[x,y]=0$. Such an algebra picks out the differential structure of the plane.\footnote{More precisely, the ring of continuous real-valued functions, and the subring of bounded functions determines the topology, since the latter's group of automorphisms is isomorphic to the group of homeomorphisms of a unique space. The subring of infinitely differentiable functions determines the differential geometry---its group of automorphisms is isomorphic to the group of diffeomorphisms. Finally, the subring of constant functions allows the definition of vector fields and hence other geometric objects. The foundation of this work (and also non-commutative geometry) is \cite{I.MM.A:43}; for an introduction see chapter 1 of \cite{J.M:01}.} Further, Robert \citet{ger:72a} pointed out that all the geometric objects involved in general relativity can be re-described as objects acting on such an algebra (for instance a derivative is clearly a special map of the algebra onto itself). John \citet[\S9.9]{ear:89b} proposed using such `Einstein algebras' to short-cut the hole argument: arguably, diffeomorphisms do not change the algebra (but see \citealt{ryn:92a}). 

Once you know that (formally speaking) ordinary geometry can be equivalently  formulated in terms of a commutative algebra of fields, it's natural to consider generalizing by choosing $[x,y]=\theta_{x,y}\neq0$, some deformation from commutativity. Can all the necessary geometric objects still be defined? Yes, as can those necessary for treating a fiber bundle over the manifold---all the mathematical apparatus of modern Lagrangian physics exists in the non-commutative algebra generalizing familiar commutative geometry.\footnote{Alain Connes is the pioneer of this work: e.g., \cite{Con:94}. See \cite{dounek:01a} and \cite{sza:03b} for good introductions.} Non-commutative versions of familiar field theories can be readily written down, so it seems it is a contingent matter whether such a theory is correct.

Such a theory simply does not contain a manifold in its formulation. The fundamental objects are the elements of the algebra, scalar `fields', identified only by their algebraic relations to the other `fields', and not by their `point-values'. Indeed, that concept has no place in the algebra. In the commutative case it turns out that the elements of the algebra can be \emph{represented} by scalar fields; since that representation involves a manifold one can ask for the value at a point, but it's important to realize that only whole fields, defined over the whole manifold, correspond to elements of the algebra. Still, in the commutative case, it's natural to take the representation as a literal description of reality and the algebra as a mere abstraction because the former corresponds to our experiences (though note that Earman is at least toying with the idea that the hole argument should drive us to the algebraic ontology). But what about the non-commutative case? The usual manifold representation doesn't exist (its coordinates commute!) so if the physics of the world were described by a NCFT, could it possibly be related to spacetime or local beables? Or do we finally have a clear case of empirical incoherence? 

No, we do not, because there is a technical strategy for solving the problem of deriving predictions that are local in phenomenal space, from the algebra. The non-commutative algebra also has a representation in an ordinary space, if one introduces a new product for scalar fields over the manifold, the `Moyal star', $\star$, according to which coordinates don't commute: $x\star y-y\star x \equiv [x,y]_\star=\theta_{x,y}$ (where this relation holds point-wise for the fields $x(p)$ etc). Then an NCFT maps into a field theory on the manifold, but with ordinary multiplication replaced by $\star$-multiplication. The usual techniques of QFT are then applied and phenomenal local beables, typically in the form of scattering events, predicted---indeed, such a calculation is exactly what one finds in physics papers on the topic. This is not to say that there are no problems in understanding the empirical content of NCFT (for instance, renormalizability depends on how the details are filled in), but it is to say that there is a technical strategy for cashing it out. 

Given this strategy, why not again take the representation to be literal? To do so would commit one to the reality of physically otiose structure, since the actual commutativity of coordinates plays no physical role; it cannot be probed by any physical process, since these only respect $\star$-multiplication. (Unsurprisingly the Moyal star is commutative at zeroth order, explaining the appearance of a commutative space in low energy phenomena. So, more carefully, commutative multiplication is otiose except insofar as it provides an approximation to physical, $\star$-multiplication.) That is, insisting on the fundamental reality of the manifold is much like insisting on the reality of an absolute standard of simultaneity in special relativity.

In short, NCFT probably represents the biggest conceptual break with the conventional idea of spacetime and hence of local beables, but it turns out that there is a rather direct way of mapping the purely algebraic structure into entities that look like local beables.

\subsection{Emergent gravity}

For greater completeness we want to acknowledge, more or less in passing, a growing number of other iconoclastic approaches to quantum gravity that threaten the fundamentality of spacetime in another way. The complexity of the situation, however, means that we will not attempt to say how they bear on the issue of empirical significance.\footnote{Bain and Crowther (this volume) address approaches in this category.}

Following the lead of Andrei \citet{sak68}, these approaches take gravity to be an intrinsically classical, large-scale phenomenon arising from the collective action of the dynamics of more fundamental, non-gravitational degrees of freedom.\footnote{For a recent survey, cf.\ \citet{bareal11}; \citet[\S3.1]{wut05} gives an accessible and short statement of the main ideas of Sakharov's groundbreaking essay.} Gravity would thus be a residual force, not an elementary aspect of nature. This motley crew of approaches takes gravity to be an ultimately thermodynamical, hydrodynamical, solid-state-physical, or computational phenomenon which emerges at the large scales or low energies that are usually accessible to us. Thus, gravity only emerges from an even more radically different physical structure or from non-gravitational physical processes or forces. For instance, it could be the case that the general-relativistic Einstein-Hilbert action can be found to some order of an effective Lagrangian of some fundamental quantum fields propagating on a fixed, but flat, background spacetime. 

Whether, and if so to what extent, however, the non-fundamentality of gravity entails the non-fundamentality of spacetime differs for each of these approaches and would have to be addressed in a separate enquiry. What ought to be emphasized here, however, is that the entailment is certainly not automatic, as approaches in this family sometimes reject the general-relativistic understanding of gravity as encoded in the geometry of spacetime. Thus, they may assume a flat and fixed Minkowskian background spacetime on which the non-gravitational fundamental interactions among elementary physical entities occurs and give rise to gravity at some appropriately larger scales. Hence, depending on the particular approach spacetime may or may not be emergent.\\

Before returning to the issue of empirical coherence, let us stress an important difference in the route back to local beables taken by approaches of types (i)-(iii), vis-\`a-vis those of types (iv)-(v). In the first three cases, where we had either a discrete structure in the form of a spatiotemporal lattice or of a fundamental causal set or else the discrete and non-local spin networks of LQG, the main aim in regaining local beables is to show how spacetime, and in particular {\em relativistic} spacetimes, re-emerge from the fundamental structure in an appropriate limit of large scales or low energies. The presumption here is that if we can recover the usual continuous spacetime, hopefully with some energy-matter content, then we have reduced the problem of identifying local beables to the classical problem of identifying observables (even if this latter is non-trivial). For approaches in the second group, string theory and NCFT, the path to local beables does not lead primarily through a recovery of a relativistic spacetime, but instead through the identification of magnitudes connected with scattering events---and the usual phenomenology of particle physics.

\section{Empirical coherence revisited}

We apologize if the reader has mental whiplash after that roller-coaster tour of several decades of intensive research in fundamental physics, and some of the deep conceptual questions that go with it. Clearly our account does not do justice to the many relevant issues raised---that would require at least a book.\footnote{We can report happily that we are currently writing such a volume.} But let's refocus on the point of this excursion for the present work. 

We have given sufficient detail to make a few points clear. First, there are grades of non-spatiality, or non-spatiotemporality: mere discreteness is not a great conceptual leap from ordinary spacetime, an algebraic ontology is, and there are cases in-between. Second, and perhaps counter-intuitively, the technical difficulty of obtaining a spacetime structure---and hence the possibility of local beables---need not increase monotonically as theories become more conceptually removed from spacetime. For instance, causal sets may preserve an important aspect of spacetime, but it doesn't permit the ready identification of local beables, while if we take dualities seriously, important aspects of spacetime are missing in string theory, and yet local beables are available. So finally, we have seen fairly convincingly that there is no general argument that the omission of spacetime from the fundamental ontology will automatically produce \emph{technical} problems in the derivation of local beables that will lead to empirical incoherence.

Still, we think that legitimate \emph{conceptual} questions about the significance of the derivation remain, questions raised particularly sharply by Maudlin. Our final goal is to address them---and in so doing demonstrate the philosophical significance of the question of how spacetime might `emerge' from a theory without spacetime.

First, Maudlin:
\begin{quote}
But one might also try instead to derive a physical structure with the form of local beables from a basic ontology that does not postulate them. This would allow the theory to make contact with evidence still at the level of local beables, but would also insist that, at a fundamental level, the local structure is not itself primitive. ... This approach turns critically on what such a derivation of something isomorphic to local structure would look like, \emph{where the derived structure deserves to be regarded as physically salient} (rather than merely mathematically definable). Until we know how to identify physically serious derivative structure, it is not clear how to implement this strategy. \citep[3161, emphasis added]{mau07}
\end{quote}

We have italicized the key phrase here. Suppose one managed to show that certain derivative quantities in a non-spatiotemporal theory took on values corresponding to the values of spatiotemporal quantities; one would have an algorithm for generating predictions about phenomenal space (especially material systems in space). According to the passage quoted, such a derivation (even if the predictions were correct) would not show that spacetime had really been derived; in addition, we have to be assured that the derived structure is `physically salient'. 

If one, say, has a theory of small things in space, and one derives from their collective behavior the behavior of large things, then there is a pretty clear sense in which the identified, collective, structure is `physically salient'---the physics of the whole depends on the physics of its spatial parts. But once we consider theories without spacetime, and perhaps lose the notion of spatial parts, this analysis cannot be sustained. Perhaps some non-spatial analogue of part is needed, which builds in `physical salience' in some way? But what would make some other sense of parthood itself physically salient? So the worry is that the demand of physical salience can never be satisfied by a theory without fundamental spacetime; in which case a merely technical derivation of putative `local beables' can never show that they can be identified with the local beables of experiment and observation; that is, the conceptual gulf between spatiotemporal phenomena and the fundamental theory is too wide to bridge---there is, after all, a general argument for empirical incoherence.

Without actually giving a specific account of `physical salience' that could encompass all the theories we've discussed (let alone all {\em possible} theories of the kind), some reflection on what work such a concept is supposed to do in science will show that Maudlin's challenge evaporates. Before explaining why, we want to briefly raise, and quickly put aside, one kind of question: `why is it that we have \emph{direct experiences} of local beables, if fundamentally there are none?' First, the conditional is surely a red herring: many (perhaps all) of the things we experience are non-fundamental (herrings and redness, for instance), so that alone can't be a problem. The usual story that we tell in the case of perceived non-fundamental entities has to do with the physical properties of events, causing changes in the environment, which propagate and cause physical changes in us. As a result \dots\ by the mind-body magic, we perceive the events. But if a theory without fundamental local beables has any chance at being correct, everything on the physical side of the usual story will have to be derivable---so the only issue remaining is how the mind-body connection is understood within the theory. But this, of course, is a question that every theory must face, so it can hardly be posed as a special challenge to those without spacetime. (Indeed, despite occasional suggestions to the contrary, it's unlikely that any details of fundamental physics will shed light on the question at all, meaning that the mind-body story would have to get worked out at a higher level.)

That kind of concern aside, there are two ways of looking at the idea of physical salience: `from below', as Maudlin seems to, taking for granted that the theory has physical salience, and asking what formal derivations of higher-level structures preserve it. But one can equally legitimately consider the idea `from above', taking for granted that the empirical realm has physical salience (which it certainly does), and asking how it can `flow down' formal derivations to give physical significance to the underlying theory.

First, starting `from below', suppose that we agree that some theory without fundamental spacetime---say a NCFT, formulated algebraically---captures physical reality: it has physical salience. And suppose that a technical derivation of local beables is found: as, for instance, in the derivation of scattering events via the $\star$-multiplication representation. Our current conception of physical salience will not imply that the derived structure inherits it from the fundamental theory, because that conception is inherently spatiotemporal: the derived structure isn't `composed' of smaller parts, it doesn't result from some `mechanical process' of more fundamental entities, it isn't some `localized part' of some more diffuse structure, and so on. But thus acknowledging that the derived structure could never inherit `physical salience' in the ways understood by our current theories, serves to highlight the fact that the very concept is theory-dependent, and hence that the argument from below is essentially question begging.

Of course there are historical precedents. To return once more to the seventeenth century, the Cartesians (sensibly) argued that only derivations involving changing configurations of geometric pieces of matter, and collisions between them were allowable---preserved `physical salience' we might say. Any other higher-level structures would have to be put down to chance, and were otherwise physically inexplicable. But they were wrong, something learned from finding theories that better match the phenomena: Newtonian gravity allows action at a distance. (Of course that was wrong too, and in (commutative) field theories we take some notion of locality to be essential; and so on.) These theories share the idea that physical salience involves spatiotemporal notions, but that's no surprise (and no argument either) because they all assume space and time! When one considers theories with spacetime in the fundamental theory, one simply has to expect that any notion of physical salience will have to shift accordingly. Whilst eschewing any conclusions about irrationality, physical salience is thus exactly the kind of philosophical concept that Thomas \cite{Kuh:62} identified as part of any theory, and which undermine any facile accounts of theory comparison. In short, we say Maudlin's challenge amounts to inappropriately holding theories without spacetime to the standards of theories that postulate it.

To be fair, the end of the quotation does leave open the possibility of developing the idea of physical salience to encompass non-spacetime theories. Clearly we do not believe that such a question can be addressed a priori, but how might it be addressed within a theory? It seems to us that it will be answered by considering what formal derivations actually produce structures that are in agreement with elements of the empirical realm. If the supposed derivation from a theory without spacetime produces quantitative agreement with the observed properties of local beables---and hence derived mathematical counterparts of local beables and their observable properties---then that is evidence both for the theory and the physical salience of the derivation. We will learn what is physically salient in the theory by learning how, in general, to make successful predictions from it. Looked at this way, Maudlin has the order of things reversed: we should not figure out what it is to be physically salient and then use the concept to validate derivations. Instead, the empirical success of the derivations is already our best guide to their salience. Thus there is no conceptual challenge to empirical coherence `from below'.

This discussion has thus already led us to look at the issues `from above'. What we want to do now is spell out more carefully how physical salience can flow down, in a way that poses no threat to empirical coherence---and clarifies the philosophical project of understanding the emergence of spacetime. David \citet[inspired by Carnap]{lew70} proposed (broadly speaking) to understand the meaning of theoretical terms in the following way. Suppose we have a new theory, which postulates and describes a new entity $\tau$---that's what we mean by saying $\ulcorner\tau\urcorner$ is a `theoretical' term. For the theory to get some purchase on the physical world as we understand it, it will also have to carry implications for things we are already familiar with; formally, this can be achieved if it contains terms that are antecedently defined---let $\ulcorner o\urcorner$ be such an `old' term in the theory. To make things semi-formal, suppose that the theory can be expressed as a sentence $T(\tau,o)$. Lewis then proposes that when scientists talk about $\tau$ they mean the \emph{unique} $x$ such that $T(x,o)$ (the open sentence obtained by replacing every occurrence of $\tau$ with $x$). That is, the meaning of $\tau$ is defined in terms of its logical place in the theory in relation to terms whose meaning we already understand (with a few modifications to generalize to theories with more terms). Of course, the theory itself asserts the existence of this entity.

There are undoubtedly problems with this story as a general view of meaning in science (it makes meaning dependent on the exact truth of a well-defined theory) but equally it undoubtedly captures something right about the way that scientists understand terms of which they only have a grasp through theory---as often happens in the cutting edge of theoretical physics, in particular.\footnote{The problems mentioned are typical for description-based theories of reference, and are exploited by Kripke, for instance, to argue for a theory of direct reference. In our view, scientists are opportunists about reference, variously using description or some mode of direct reference, depending what is available and best suited to their current purposes.} Now, in the case at hand, the situation is a little different from that considered by Lewis, because something \emph{derived} will, by definition, not appear in the theory, and so the corresponding old term will not appear in the theory sentence. Instead, suppose we have a theory, $T(\tau_1,\tau_2,\dots,\tau_n)$, of some non-spatiotemporal entities, $\tau_1,\tau_2,\dots,\tau_n$, and a demonstration that, given suitable idealizations, some formal structure can be derived in which certain variables are functionally related just as phenomenal---`old'---spacetime quantities. (Actually, we suspect that this sort of situation is the norm in physics, rather than the formally cleaner case considered by Lewis.) But the general scheme still works: the $\tau$s are defined to be the unique collection of things satisfying the theory, such that \emph{the structure in question veridically represents the spatiotemporal quantities}. So, by definition, if the $\tau$s exist, there is no further question of whether spacetime emerges from them, since they just are (in part) the things from which spacetime emerges. 

One is rightly suspicious of attempts to define away problems, and an aspect of Maudlin's concern still arises---but in a form in which it can be addressed. Now the question is whether there are unique things answering the theoretical description of $\tau_1,\tau_2,\dots,\tau_n$ such that the identified emergent structure is phenomenal spacetime. Put existentially this way, the problem is clearly one of scientific realism: do we have adequate evidence for such entities or not? This is hardly the place to enter a general debate, so the only question is whether there would be \emph{special} reasons for doubting the truth of the theory-sans-spacetime at stake, given empirical evidence. In the first place, we do not think that the passage that we quoted clearly identifies any. Second, we believe that any such special worries can only be sensibly investigated through a study of particular theories purporting to describe emergent spacetime, not a priori, via the most general considerations. So here we will simply note (as many have before) that the further one gets from the phenomena, the more tenuous the chain of evidence becomes, and otherwise table the challenge.

So Maudlin's challenge is defused. But the points that we have just made have a far broader significance than we have so far seen. The program of interpreting a theory `from above', of explicating the empirical significance of a theory, is both `philosophical' in the sense that it requires the analysis of concepts, and crucial to every previous advance in fundamental physics. While we don't agree with every conclusion that he draws, such is a basic lesson of Michael \citet{Fri:01}; Robert \citet{DiS:06} comprehensively illustrates the point in the history of spacetime physics. Moreover, we agree with them that the historical record shows that such work is not a matter of waiting until an uninterpreted formalism is developed (as the logical positivists assumed when trying to make sense of this aspect of science). Rather it is work that is and must be carried out in parallel with formal developments. As such, it must be pursued by the study of theory fragments, toy models and false theories capturing some promising ideas, asking how empirical spacetime relates to them---in effect, studying how spacetime emerges from theories without fundamental spacetime. For that reason we believe that philosophers are well placed to assist in the development of the next physics, so that the significance of studying theories of quantum gravity goes far beyond the novel metaphysics that they hint at. We are very pleased that this essay is appearing in a volume of work taking the first steps in this direction. We strongly encourage all philosophers of physics to pay special attention.

\bibliographystyle{plainnat}
\bibliography{biblio}

\end{document}